\documentclass[preprint]{aastex}
\usepackage{epsfig,lscape}
\slugcomment{KSUPT-03/2 \hspace{0.5truecm} February 2003}

\newcommand{\om}{\Omega_{{\rm M}0}}
\newcommand{\kmsmpc}{{\rm \, km\, s}^{-1}{\rm Mpc}^{-1}}

\begin{document}

\title{Median Statistics and the Mass Density of the Universe}

\author{Gang Chen and Bharat Ratra}

\affil{Department of Physics, Kansas State University, 116 Cardwell Hall, 
Manhattan, KS 66506.}

\begin{abstract}
We use weighted mean and median statistics techniques to combine individual
estimates of $\om$, the present mean mass density in non-relativistic 
matter, and determine the observed values and ranges of $\om$ from different
combinations of data. The derived
weighted mean $\om$ values are not good representatives of the individual
measurements, under the assumptions of Gaussianity and negligible correlation
between the individual measurements. This could mean that some observational 
error bars are under-estimated. Discarding the most discrepant $\sim$ 5 \% of
the measurements generally alleviates but does not completely resolve this
problem. While the results derived from the different combinations of data 
are not identical, they are mostly consistent, and a reasonable summary of
the median statistics analyses is $0.2 \la \om \la 0.35$ at two standard
deviations.
\end{abstract}


\section{Introduction} 

Oftentimes it is useful to combine results from many different measurements 
of a quantity and derive a more accurate estimate of that quantity. Thus
Gott et al.~(2001) study a collection of all available pre-mid-1999
estimates of the present value of Hubble's constant, $H_0$, and derive
\begin{equation}
  H_0 = 100 h \kmsmpc = 67 \pm 7 \kmsmpc ,
\end{equation}
at two standard deviations\footnote{
Where the first equation defines the dimensionless parameter $h$ and we
halve the 2 $\sigma$ error bar of Gott et al.~(2001) to get a 1 $\sigma$
error bar for our computations.},
a significantly more constraining estimate of $H_0$ than is provided by 
any single measurement.

Similar meta-analysis techniques have been used to determine binned
multipole-space cosmic microwave background (CMB) anisotropy power 
spectra by combining many different CMB anisotropy measurements (see,
e.g., Podariu et al.~2001; Miller et al.~2002; Page 2002; Wang et al.~2002;
Mukherjee \& Wang 2003), 
and to derive constraints on cosmological-model-parameters from
combined CMB anisotropy data sets (see, e.g., \"Odman et al.~2002;
Mukherjee et al.~2002; Douspis et al.~2002; Wang et al.~2002).

The more widely used weighted mean technique, discussed in Podariu 
et al.~(2001), assumes Gaussian errors.\footnote{
A number of quantities of interest, e.g., the CMB anisotropy spectrum,
are commonly thought to have been generated by quantum fluctuations in
a weakly coupled field during an early epoch of inflation and are thus
realizations of spatially stationary Gaussian random processes (see, 
e.g., Ratra 1985; Fischler, Ratra, \& Susskind 1985). Measurements 
appear to be consistent with this Gaussianity assumption --- for discussions 
of the Gaussianity of the smaller-scale CMB anisotropy see, e.g., Park et 
al.~(2001), Shandarin et al.~(2002), Santos et al.~(2002), and Polenta
et al.~(2002) --- and so in cases where the experimental noise is Gaussian
it is fair to use the weighted mean technique.}
Thus in this case one may compute a goodness-of-fit parameter, and the 
number of standard deviations, $N_\sigma$, this parameter deviates 
from what is expected (Podariu et al.~2001). A large value of $N_\sigma$
could indicate the presence of unaccounted for systematic uncertainties, 
the breakdown of the Gaussian assumption, or the presence of significant
correlations between the individual measurements used.   
 
The other technique we use, that based on median statistics (Gott et 
al.~2001)\footnote{
See Avelino, Martins, \& Pinto (2002) for a recent application of the
Gott et al.~(2001) median statistics technique.}, 
does not assume that the measurement errors are Gaussian, or even that 
the magnitude of these errors are known. It assumes only that the 
measurements are independent and free of systematic errors. It is hence 
not possible to estimate the goodness of fit in the median statistics
case. However, since the median statistics technique is based on fewer
assumptions than the weighted mean technique, median statistics results 
are more robust, but still --- as Gott et al.~(2001) show --- almost as
constraining as weighted mean results.

In this paper we apply both these techniques to collections of estimates
of $\om$, the present value of the mean mass density of non-relativistic 
matter in the universe. A robust and tight estimate of $\om$ is of great 
interest. Current indications are that $\om$ is small and we live in a 
low-density universe (see Peebles \& Ratra 2003 for a review). This, in
conjunction with recent CMB anisotropy measurements which suggest that 
the curvature of spatial hypersurfaces is small, indicates that a 
dark energy dominated spatially-flat universe (see, e.g., Peebles 1984; 
Peebles \& Ratra 1988, 2003; Steinhardt 1999; Sahni \& Starobinsky 2000;
Carroll 2001; Padmanabhan 2002) is observationally favored over a 
spatially open model with insignificant dark energy density (see, e.g., 
Gott 1982; Ratra \& Peebles 1995). To strengthen this conclusion it 
would be helpful to have in hand a more robust and tight estimate of $\om$
than is available from any single measurement.\footnote{
Of course, comparing the predictions of dark energy dominated models to
observational measurements is another way to check for the presence of
dark energy. In the near future neoclassical cosmological tests that
hold significant promise include those based on CMB anisotropy (see,
e.g., Brax, Martin, \& Riazuelo 2000; Amendola et al.~2002), gravitational
lensing (see, e.g., Ratra \& Quillen 1992; Waga \& Frieman 2000; Chae 
et al.~2002), Type Ia supernova redshift-apparent magnitude (see, e.g.,
Podariu \& Ratra 2000; Waga \& Frieman 2000; Leibundgut 2001), 
redshift-counts (see, e.g., Huterer \& Turner 2001; Podariu \& Ratra 
2001; Levine, Schulz, \& White 2002), and redshift-angular size (see,
e.g., Zhu \& Fujimoto 2002; Chen \& Ratra 2003; Podariu et al.~2003)
data.}

While it would be useful to focus on measurements of $\om$ that are
independent of cosmological model, we have been able to locate only 30 
such recent smaller-scale estimates of $\om$. To reduce
statistical uncertainty it is desirable to have a greater number of
$\om$ estimates. We hence also consider recent $\om$ estimates derived 
assuming either a spatially-flat model with a cosmological constant
$\Lambda$ or an open model with no $\Lambda$. 

The $\om$ measurements we focus on are listed and discussed in $\S$ 2.
Results are presented and discussed in $\S$ 3. We conclude in $\S$ 4. 

\section{$\om$ Measurements}

Tables 1---3 list the values and errors bars of $\om$ for the measurements 
we consider. Table 1 lists values determined in a manner that is independent 
of cosmological model, while Tables 2 and 3 list values derived assuming a 
spatially flat $\Lambda$ dominated model, and a spatially open model with
no $\Lambda$, respectively. In general, in these Tables, we include quoted 
systematic errors in quadrature when determining the total error bar and
assume a Gaussian distribution when determining 1 $\sigma$ errors (if these
are not given). In what follows we briefly describe how we determine the
$\om$ values and error bars given in these Tables.

\subsection{Redshift Distortion Factor}

There are many measurements of the redshift distortion factor
$\beta=\om^{0.6}/b$, where $\om^{0.6}$ is a reasonably accurate approximation 
of the velocity function evaluated at zero redshift $f(z = 0)$ (Peebles 
1993, $\S$ 13) and $b$ is the bias factor for the tracer used, defined in 
terms of the ratio of the fractional density perturbations, 
$b = \delta_{\rm trace}/\delta_{\rm mass}$, where $\delta_{\rm trace}$ is
the fractional number density perturbation. 

To determine $\om$ from $\beta$ we need to know the bias factor. In this 
paper, we use, for optical galaxies (Verde et al.~2002; Lahav et al.~2002; 
Peacock et al.~2002)\footnote{
We average the values given in these papers to get the $b_O$ value quoted 
here.}
\begin{equation}
  b_O = 1.0 \pm 0.1 , 
\end{equation}
and for infrared galaxies and clusters of galaxies (Plionis et al.~2000),
\begin{equation}
  b_I=b_O/(1.21 \pm 0.06) , \qquad  b_C=b_I (4.3 \pm 0.8) ,
\end{equation}
all at one standard deviation.

Typically, $\beta$ is measured from density-density comparisons (D-D in the
Tables) or velocity-velocity comparisons (V-V in the Tables), or through the 
distortion effect of peculiar velocities on redshift surveys. By using the 
somewhat related least-action principle, Susperregi (2001) is able to 
independently determine $\om$ and $b$, and we quote his value of $\om$ in 
Table 1.

\subsection{Power Spectrum}

A commonly used simple analytic fit to the observed power spectrum of 
cosmological mass fluctuations is the CDM spectrum (see, e.g., Peacock 
1999, $\S$ 16.8). In this approximation, the shape and amplitude 
of the mass power spectrum depends on two parameters: the shape parameter 
$\Gamma$, and $\sigma_8$, the rms fractional mass density variation averaged 
over 8$h^{-1}$ Mpc spheres. Nowadays, the shape parameter is usually 
approximated by (Sugiyama 1995),
\begin{equation}
  \Gamma = \om h{}e^{ -{}\Omega_B{}(1 + \sqrt{2h} / \om) }
\end{equation}
where $\Omega_B$ is a measure of the  present mean mass density 
in baryonic matter. Occasionally however the shape parameter is 
still defined through $\Gamma=\om h$. 

To extract a value and error bars for $\om$ from
a measurement of $\Gamma$ we need the value of $h$ --- we use 
eq.~(1) for this --- and an estimate of the baryonic mass density
parameter $\Omega_B$, for which we use
\begin{equation}
  \Omega_B h^2 = 0.014 \pm 0.004 , 
\end{equation}
at 1 $\sigma$, derived by averaging the two extreme values quoted 
in $\S$ IV.B.2 of Peebles \& Ratra (2003).

\subsection{Velocity Correlation}

By assuming a shape for the power spectrum, the velocity correlation method 
provides a constraint on a function of $\sigma_8$ and $\om^{0.6}$ by 
comparing the observed velocity correlation to that predicted from the 
power spectrum. Given an estimate of $\sigma_8$, for which we use\footnote{
Here the rms factional mass density variation averaged over 8$h^{-1}$ Mpc 
spheres $\sigma_8 = \sigma_{8{\rm trace}}/b_{\rm trace}$, where 
$\sigma_{8{\rm trace}}$ is the corresponding rms number density variation
and $b_{\rm trace}$ is the bias factor for the tracer used. We average the 
values given for $\sigma_8$ in eq.~(37) of Hamilton \& Tegmark (2002) and 
eq.~(12) of Szalay et al.~(2001).}
\begin{equation}
 \sigma_8 = 0.94 \pm 0.11 ,
\end{equation}
at 1 $\sigma$, we can determine $\om$. Juszkiewicz et al.~(2000) consider a 
variant of this method based on relative velocities of galaxies.

\subsection{Gas Mass Fraction}

Assuming the baryonic mass fraction in galaxy clusters is an accurate 
representation of that of the universe, and given $\Omega_B$, the baryon 
mass fraction $f_b$ provides an estimate of $\om$,
\begin{equation}
  \om = \Omega_B/f_b .
\end{equation}
The related gas mass fraction of galaxy clusters, $f_g$, is what is 
measured, and it provides the estimate
\begin{equation}
  \om={{\Omega_B}\over{f_g (1+0.19 h^{0.5})}} ,
\end{equation}
where we use $h$ and $\Omega_B$ from eqs.~(1) and (5).

\subsection{Mass to Light Ratios}

This method assumes that the mass-to-light ratios of galaxy clusters are 
accurate representatives of that of the whole universe. Table 1 lists the 
Carlberg et al.~(1997a) and Bahcall et al.~(2000) $\om$ values and 1 
$\sigma$ ranges. We also show results derived from the Hradecky et al.~(2000)
data; to determine the $\om$ central value and 1 $\sigma$ range, we 
compute the weighted mean and error bar of $M/L_V$ using Table 5 of their 
paper and use the Efstathiou, Ellis, \& Peterson (1988) estimate 
$\langle L \rangle \approx (2 \pm 0.7) \times 10^8 h L_\odot$ Mpc$^{-3}$. 

\subsection{Cosmological-Model-Dependent Estimates}

Tables 2 and 3 list $\om$ values determined assuming a flat-$\Lambda$ 
and an open cosmological model, respectively. Such model-dependent estimates 
are becoming more common. Weak lensing (WL) measurements have recently 
begun to provide interesting constraints on a function of $\om$ and
$\sigma_8$, while improving galaxy cluster number density measurements 
(both at the present epoch and as a function of redshift) constrain
a related function. Using the estimate for $\sigma_8$ given in eq.~(6)
we may use these constraints as measurements of $\om$.  
 
Table 2 also lists estimates of $\om$ from various other methods, including
cosmic microwave background (CMB) anisotropy measurements, the angular size
versus redshift ($\theta-z$) test, strong gravitational lensing, and the
supernova apparent magnitude versus redshift test. 

\section{Methods and Results}

Table 1 lists 30 ``model-independent" measurements of $\om$, while Tables 2
and 3 show 28 and 14 results determined assuming a flat-$\Lambda$ and an open 
model, respectively. In addition to these three data sets, we also consider
two additional combination data sets: combinations of the 30 model-independent 
results with the flat-$\Lambda$ model values and with the open model values.

For each of these five data sets, we compute the weighted mean of $\om$ 
and the associated error estimate as follows (see, e.g., Podariu et 
al.~2001). The standard expression for the weighted  mean is
\begin{equation}
  \om={\sum_{i=1}^N (\om)_i / \sigma_i^2 \over \sum_{i=1}^N 1/\sigma_i{}^2} ,
\end{equation}
where $i=1,2, \dots N$ indexes the $N$ measurements in the data set, with
central values $(\om)_i$ and errors $\sigma_i$. The (internal) error estimate 
for each data set is
\begin{equation}
  \sigma=\left(\sum_{i=1}^N 1/\sigma_i{}^2\right)^{-1/2} .
\end{equation}
The goodness of fit parameter is
\begin{equation}
  \chi^2={1 \over N-1}\sum_{i=1}^N {\left((\om)_i - \om \right)^2 \over 
  \sigma_i{}^2} = \sum_{i=1}^N \chi^2_i ,
\end{equation}
where the last equation defines $\chi^2_i$, the ``reduced $\chi^2$" 
contribution from each measurement. 
Since the weighted mean technique assumes Gaussian errors, $\chi$ has 
expected value unity with error $1/\sqrt{2(N - 1)}$, so the number 
of standard deviations that $\chi$ deviates from unity is
\begin{equation}
  N_\sigma = |\chi - 1| \sqrt{2(N-1)} .
\end{equation}
A large value of $N_\sigma$ could indicate the presence of unaccounted for 
systematic errors, the invalidity of the Gaussian assumption, or the 
presence of significant correlations between the measurements.

We also analyze each of the five data sets using median statistics (see, e.g., 
Gott et al.~2001; Podariu et al.~2001). For each data set, we construct  
the distribution for the true median $\om$ value using the binomial 
theorem method of eq.~(1) of Gott et al.~(2001).\footnote{
For example, if there were two measurements of a given quantity, $x_1$
and $x_2$ ($x_1 < x_2 $), then there is a 25 \% probability that the true
median lies below $x_1$, a 50 \% probability that it lies between $x_1$ and
$x_2$, and a 25 \% probability that it lies above $x_2$.}
Since $\om$ is positive, following Gott et al.~(2001), we integrate over 
this distribution with a logarithmic prior between data points to 
determine confidence intervals for $\om$.\footnote{
Lower and upper confidence levels at 1 and 2 $\sigma$ significance are 
determined such that the probabilities inside the 1 and 2 $\sigma$ ranges 
are divided in half by the median value.}

Table 4 shows the results for the weighted mean and median statistics
analyses. The upper half of the table shows results derived using all 
measurements. The weighted mean technique results in tighter constraints
on $\om$, while the median statistics constraints are weaker. This 
result (Gott et al.~2001; Podariu et al.~2001) is reinforced by the 
large $N_\sigma$ values in the upper half of Table 4. For Gaussian 
distributed errors, $N_\sigma$ is a measure of how well
the weighted mean and derived error bar represent the measurements 
considered. $N_\sigma$ is greater than 2 in all cases, i.e., $\chi$ is 
more than 2 $\sigma$ away from what is expected for Gaussian distributed 
errors. This most likely means that one (or more) of the measurements 
has an underestimated error bar.\footnote{
Given the available evidence, it is reasonable to assume that the Gaussianity
assumption is not invalid. In addition, we have attempted to select only
those measurements that are not strongly correlated. Thus the large
$N_\sigma$ values we find are most likely a consequence of one (or more)
measurements that have underestimated error bars.} 
Since median statistics do not make use 
of the measurement error bars, the median statistics results are likely more 
reliable than the weighted mean results.

To examine the issue of large $N_\sigma$ values, we proceed as follows.
For each measurement in each of the five data sets, we compute $\chi^2_i$,
eq. (11). We then discard the $\sim$ 5 \% most discrepant (largest $\chi^2_i$)
measurements from each data set and so generate five culled data sets
of ``good" measurements. For the model-independent data set, we drop 
only one measurement, that from the least-action principle method 
(Susperregi 2001), which 
has a small error bar and $\chi^2_i = 0.99$. For the flat-$\Lambda$ data set 
we have to drop two measurements: the Zaroubi et al. (1997) $v$-correlation 
result which has a large $\om$ and $\chi^2_i = 0.29$, and the Allen et al. 
(2002b) cluster number density measurement which has a small $\om$
and $\chi^2_i = 0.27$. For the open model data set we drop the Hamana et al.  
(2002) weak lensing measurement which has a small $\om$ and $\chi^2_i = 1.3$. 
These measurements are also the most discrepant ones in the two combination 
data sets, so we drop them again when generating the culled combination 
data sets. Of course, not all large $N_\sigma$ values are reduced to unity 
by this culling (although they can be by further culling): we are only
discarding the most ``discrepant" measurements to investigate the stability
(robustness) of the constraints on $\om$.    

Results from the analyses of the culled data sets are shown in the lower
half of Table 4. For the model-independent, flat-$\Lambda$, and their 
combination data sets, the weighted mean and median statistics error
bars are in better accord, and $N_\sigma$ are of order unity. For the 
open model data sets $N_\sigma$ are smaller now, but still significantly 
large than unity, indicating perhaps that the error bars on one or more
of the remaining measurements are underestimated.

Focussing on the median statistics 2 $\sigma$ ranges for the culled 
data sets (the lower half of the last column of Table 4), if we exclude 
the open model results for the reason mentioned above, a reasonable 
summary is
\begin{equation}
   0.2 \la \om \la 0.35
\end{equation}
at two standard deviations, with central value at
$\om \sim 0.25 - 0.3$. It is reassuring that these summary values are
in agreement with other recent estimates (see, e.g., Peebles \& Ratra
2003).

\section{Conclusion}

We have determined a preliminary estimate of the mean mass density in
nonrelativistic matter, from median statistics analyses of various 
collections of measurements. The results of our meta-analysis estimate
of $\om$ appear very reasonable. More high quality data, especially
``model-independent" data, should allow for significantly more 
constraining limits on $\om$.
   
\bigskip

We acknowledge helpful discussions with R. Gott, J. Peebles, S. Podariu, and 
U. Seljak, and support from NSF CAREER grant AST-9875031.

\clearpage

\begin{deluxetable}{llccc}
\tablecaption{$\om$ Values that are Independent of Cosmological Model}
\tablewidth{0pt}
\tablehead{
\colhead{Method} &
\colhead{Data Set} &
\colhead{$\om$} &
\colhead{$\om$ (1 $\sigma$ range)} &
\colhead{Reference}
          }
\startdata
$z$-distortion  & 2dFGRS		& 0.24  & 0.16--0.32   	&Peacock et al.~(2002)\\
$z$-distortion	& $IRAS$PSCz		& 0.15	& 0.07--0.23	&Taylor et al.~(2000)\\
V-V		& ORS/SBF		& 0.13	& 0.08--0.18	&Blakeslee et al.~(2000)\\
V-V		& $IRAS$/SBF		& 0.19	& 0.12--0.26	&Blakeslee et al.~(2000)\\
V-V		& SFI/$IRAS$PSCz	& 0.17	& 0.13--0.21	&Branchini et al.~(2001)\\
V-V		& MarkIII/$IRAS$	& 0.23	& 0.17--0.29	&Willick \& Strauss (1998)\\
V-V		& ENEAR/$IRAS$PSCz	& 0.23	& 0.14--0.22	&Nusser et al.~(2001)\\
D-D		& MarkIII/Optical	& 0.60	& 0.33--0.87	&Hudson et al.~(1995)\\
D-D		& MarkIII/$IRAS$1.2Jy	& 0.60	& 0.42--0.78	&Sigad et al.~(1998)\\
D-D/V-V		& Abell,ACO/MarkIII	& 0.66	& 0.20--1.1	&Branchini et al.~(2000)\\
dipole		& XBACs			& 0.76	& 0.37--1.2	&Plionis \& Kolokotronis (1998)\\
LAP	& PSCz,ORS,MarkIII,SFI	& 0.37	& 0.36--0.38	&Susperregi (2001)\\
$\Gamma$	& SDSS			& 0.33	& 0.26--0.40	&Szalay et al.~(2001)\\
$\Gamma$	& REFLEX		& 0.34	& 0.25--0.42	&Schuecker et al.~(2001)\\
$\Gamma$	& 2dFQSO		& 0.20	& 0.04--0.36	&Hoyle et al.~(2002)\\
$\Gamma$	& APM			& 0.25	& 0.11--0.39	&Efstathiou \& Moody (2001)\\
$\Gamma$	& 2dFGRS		& 0.30	& 0.25--0.35	&Percival et al.~(2001)\\
$\Gamma$	& LCRS			& 0.24	& 0.09--0.39	&Matsubara et al.~(2000)\\
$v$-correlation	& ENEAR			& 0.36	& 0.24--0.48	&Borgani et al.~(2000)\\
$v$-correlation	& MarkIII		& 0.35	& 0.18--0.52\tablenotemark{a}	
	&Juszkiewicz et al.~(2000)\\
$f_g$		& SZ data	        & 0.20	& 0.14--0.26	&Grego et al.~(2001)\\
$f_g$		& RXJ2228+2037		& 0.12	& 0.07--0.16	&Pointecouteau et al.~(2002)\\
$f_g$		& $ROSAT$PSPC,$ASCA$	& 0.25	& 0.16--0.33	&Ettori \& Fabian (1999)\\
$f_g$		& $Chandra$		& 0.22	& 0.15--0.29	&Allen et al.~(2002a)\\
$f_g$		& $BeppoSAX,Chandra$	& 0.28\tablenotemark{b}	& 0.20--0.36\tablenotemark{b}	
	&Ettori et al.~(2002)\\
$f_b$		& $ROSAT$PSPC		& 0.44\tablenotemark{c}	& 0.29--0.59\tablenotemark{c}	
	&Sadat \& Blanchard (2001)\\
$f_b$		& $ROSAT$,$Ginga$,$ASCA$	& 0.33	& 0.23--0.42	&Roussel et al.~(2000)\\
M/L		&			& 0.19	& 0.12--0.26	&Carlberg et al.~(1997a)\\
M/L		&			& 0.16	& 0.10--0.22	&Bahcall et al.~(2000)\\
M/L		&			& 0.15\tablenotemark{d}	& 0.094--0.21\tablenotemark{d}	
	&Hradecky et al.~(2000)\\
\enddata

\tablenotetext{a}{From their Figure 2, by allowing $\sigma_8$ to vary between 0.83 and 1.05, as given in our eq.~(6).}
\tablenotetext{b}{From the weighted mean and the external error of $f_{gas}$ ($\Delta$=500) values in their Table 1.}
\tablenotetext{c}{The 1 $\sigma$ range of $f_b$ is from the first rows of 
their Tables 4 and 6, and the central value is the mean of these 1 $\sigma$
values.}
\tablenotetext{d}{See the discussion in $\S$ 2.5.}

\end{deluxetable}

\begin{deluxetable}{llccc}
\tablecaption{$\om$ Values Determined Assuming a Flat-$\Lambda$ Cosmological Model}
\tablewidth{0pt}
\tablehead{
\colhead{Method} &
\colhead{Data Set} &
\colhead{$\om$} &
\colhead{$\om$ (1 $\sigma$ range)} &
\colhead{Reference}
          }
\startdata
$v$-correlation	&MarkIII		& 0.90	& 0.66--1.1	&Zaroubi et al.~(1997)\\
WL              & VIRMOS-DESCART	& 0.30	& 0.21--0.39	&Van Waerbeke et al.~(2002)\\  
WL		& Keck,WHT		& 0.31	& 0.23--0.39	&Bacon et al.~(2002)\\
WL		& COMBO-17		& 0.48	& 0.31--0.65	&Brown et al.~(2002)\\
WL		& Suprime-Cam		& 0.15	& 0.06--0.24	&Hamana et al.~(2002)\\
WL		& RCS			& 0.26	& 0.18--0.34	&Hoekstra et al.~(2002)\\
WL		& CTIO			& 0.18	& 0.13--0.23	&Jarvis et al.~(2002)\\
WL		& MDS			& 0.30	& 0.17--0.43	&Refregier et al.~(2002)\\
cluster	        &			& 0.33	& 0.22--0.44	&Viana \& Liddle (1999)\\
cluster	        &			& 0.09	& 0.04--0.14	&Allen et al.~(2002b)\\
cluster	        &			& 0.44	& 0.32--0.56	&Henry (2000)\\
cluster 	&REFLEX			& 0.34	& 0.25--0.44	&Schuecker et al.~(2002)\\
cluster	        &			& 0.18	& 0.045--0.32	&Seljak (2002)\\
cluster	        &ROSAT,ASCA		& 0.34	& 0.26--0.42	&Pierpaoli et al.~(2001)\\
cluster	        &EMSS,RASS		& 0.27	& 0.13--0.41	&Donahue \& Voit (1999)\\
cluster	        &HIFLUGCS		& 0.12	& 0.08--0.16	&Reiprich \& B\"ohringer (2002)\\
cluster	        &SDSS			& 0.18	& 0.13--0.23	&Bahcall et al.~(2003)\\
cluster	        &RDCS			& 0.35	& 0.23--0.47	&Borgani et al.~(2001)\\
cluster	        &CNOC			& 0.40	& 0.27--0.53	&Carlberg et al.~(1997b)\\
cluster	        &			& 0.57	& 0.23--0.91	&Bahcall \& Fan (1998)\\
cluster	        &			& 0.26	& 0.17--0.35	&Wu (2001)\\
cluster	        &			& 0.87	& 0.57--1.2	&Blanchard et al.~(2000)\\
cluster	        &HIFLUGCS,$Chandra$	& 0.26	& 0.14--0.38	&Vikhlinin et al.~(2002)\\
CMB anisotropy  &			& 0.38	& 0.20--0.56	&Percival et al.~(2002)\\
$\theta$-$z$      &radio galaxies	        & 0.10	& 0.00--0.35	&Guerra et al.~(2000)\\
power spectrum	&Ly$\alpha$ forest	& 0.25\tablenotemark{a}	& 0.00--0.71\tablenotemark{a}	
	&Croft et al.~(2002)\\
strong lensing	&CLASS			& 0.31	& 0.08--0.54	&Chae et al.~(2002)\\
magnitude-$z$	&supernova		& 0.28	& 0.18--0.38	&Perlmutter et al.~(1999)\\
\enddata

\tablenotetext{a}{From their eq.~(25), assuming the spectral index $n$=1 and using the values for $\Omega_B$ and $h$ given in our $\S$ 2.}

\end{deluxetable}

\begin{deluxetable}{llccc}
\tablecaption{$\om$ Values Determined Assuming an Open Cosmological Model}
\tablewidth{0pt}
\tablehead{
\colhead{Method} &
\colhead{Data Set} &
\colhead{$\om$} &
\colhead{$\om$ (1 $\sigma$ range)} &
\colhead{Reference}
          }
\startdata
$v$-correlation	&MarkIII		& 0.90	& 0.66--1.1	&Zaroubi et al.~(1997)\\
WL              & VIRMOS-DESCART	& 0.27	& 0.21--0.33	&Van Waerbeke et al.~(2002)\\
WL		& Suprime-Cam		& 0.04	& 0.00--0.08	&Hamana et al.~(2002)\\
WL		& RCS			& 0.26	& 0.18--0.34	&Hoekstra et al.~(2002)\\
WL		& FORS1			& 0.37	& 0.25--0.49	&Maoli et al.~(2001)\\
cluster	        &			& 0.22	& 0.12--0.32	&Viana \& Liddle (1999)\\
cluster	        &			& 0.49	& 0.37--0.61	&Henry (2000)\\
cluster	        &EMSS,RASS		& 0.45	& 0.31--0.59	&Donahue \& Voit (1999)\\
cluster	        &HIFLUGCS		& 0.12	& 0.08--0.16	&Reiprich \& B\"ohringer (2002)\\
cluster	        &CNOC			& 0.40	& 0.27--0.53	&Carlberg et al.~(1997b)\\
cluster	        &			& 0.51	& 0.14--0.88	&Bahcall \& Fan (1998)\\
cluster	        &			& 0.18	& 0.08--0.28	&Wu (2001)\\
cluster	        &			& 0.92	& 0.69--1.2	&Blanchard et al.~(2000)\\
cluster	        &HIFLUGCS,$Chandra$	& 0.48	& 0.40--0.56	&Vikhlinin et al.~(2002)\\
\enddata
\end{deluxetable}

\begin{landscape}
\begin{deluxetable}{lcccccccc}
\tablecaption{Weighted Mean and Median Statistics Results\tablenotemark{a}}
\tablewidth{0pt}
\tablehead{
\colhead{Data Set} &
\colhead{$N$\tablenotemark{b}} &
\colhead{$\om^{\rm WM}$} &
\colhead{$\om^{\rm WM}$(1 $\sigma$ range)} &
\colhead{$\om^{\rm WM}$(2 $\sigma$ range)} &
\colhead{$N_\sigma$}\tablenotemark{c} &
\colhead{$\om^{\rm MS}$} &
\colhead{$\om^{\rm MS}$(1 $\sigma$ range)} &
\colhead{$\om^{\rm MS}$(2 $\sigma$ range)}
	}
\startdata
\cutinhead{All Measurements}
Model-independent & 30	& 0.32 & 0.31--0.32 & 0.30--0.33 & 7.9 & 0.24 & 0.23--0.28 & 0.20--0.33\\
Flat-$\Lambda$    & 28	& 0.22 & 0.21--0.24 & 0.19--0.26 & 2.4 & 0.30 & 0.27--0.31 & 0.26--0.34\\
Open              & 14	& 0.20 & 0.18--0.22 & 0.16--0.25 & 6.4 & 0.38 & 0.26--0.45 & 0.18--0.49\\
Flat-$\Lambda$ \& Model-ind. & 58 & 0.30 & 0.29--0.30 & 0.28--0.31 & 9.0 & 0.28 & 0.25--0.30 & 0.24--0.32\\
Open \& Model-ind.& 44	& 0.30 & 0.29--0.31 & 0.29--0.32 & 11. & 0.26 & 0.24--0.31 & 0.23--0.35\\
\cutinhead{``Good" Measurements Only}
Model-independent & 29	& 0.21 & 0.20--0.23 & 0.18--0.24 & 0.98 & 0.24 & 0.23--0.26 & 0.20--0.32\\
Flat-$\Lambda$    & 26  & 0.24 & 0.22--0.25 & 0.20--0.27 & 0.92 & 0.30 & 0.27--0.31 & 0.26--0.34\\
Open              & 13  & 0.26 & 0.24--0.29 & 0.21--0.31 & 4.4  & 0.40 & 0.27--0.46 & 0.22--0.50\\
Flat-$\Lambda$ \& Model-ind. & 55 & 0.22 & 0.21--0.23 & 0.20--0.24 & 1.4  & 0.27 & 0.25--0.30 & 0.24--0.31\\
Open \& Model-ind.& 42 & 0.22 & 0.21--0.24 & 0.20--0.25 & 3.8  & 0.26 & 0.24--0.31 & 0.23--0.34\\
\enddata
\tablenotetext{a}{Superscripts WM and MS indicate weighted mean and median
statistics results, respectively.}
\tablenotetext{b}{Number of measurements in the data set.}
\tablenotetext{c}{Eq.~(12).}
\end{deluxetable}
\end{landscape}

\end{document}